# Design and Evaluation of Ultra-Fast 8-bit Approximate Multipliers Using Novel Multicolumn Inexact Compressors

Fereshteh Karimi [1], Reza Faghih Mirzaee [2*], Ali Fakeri-Tabrizi [3], Arman Roohi [4]

[1] Department of Computer Engineering, North Tehran Branch, Islamic Azad University, Tehran, Iran
[2] Department of Computer Engineering, Shahr-e-Qods Branch, Islamic Azad University, Tehran, Iran
[3] Independent Contributor, San Francisco Bay, California, USA
[4] School of Computing, University of Nebraska-Lincoln, NE, USA
* Corresponding Author's EMail: r.f.mirzaee@qodsiau.ac.ir

**Abstract**: A multiplier, as a key component in many different applications, is a time-consuming, energy-intensive computation block. Approximate computing is a practical design paradigm that attempts to improve hardware efficacy while keeping computation quality satisfactory. A novel multicolumn 3,3:2 inexact compressor is presented in this paper. It takes three partial products from two adjacent columns each for rapid partial product reduction. The proposed inexact compressor and its derivates enable us to design a high-speed approximate multiplier. Then, another ultra-fast, high-efficient approximate multiplier is achieved utilizing a systematic truncation strategy. The proposed multipliers accumulate partial products in only two stages, one fewer stage than other approximate multipliers in the literature. Implementation results by Synopsys Design Compiler and 45 nm technology node demonstrates nearly 11.11% higher speed for the second proposed design over the fastest existing approximate multiplier. Furthermore, the new approximate multipliers are applied to the image processing application of image sharpening, and their performance in this application is highly satisfactory. It is shown in this paper that the error pattern of an approximate multiplier, in addition to the mean error distance and error rate, has a direct effect on the outcomes of the image processing application.

- **Funding**: No funding was received for conducting this study
- **Conflict of Interest**: The authors have no conflicts of interest to disclose
- **Ethics/Permissions**: Not applicable

## I. Introduction

Today's modern systems and large-scale applications require a vast number of resources, including energy-storage modules, memories, and computational units [1]. Additionally, high-performance, energy-efficient electronic circuits are an absolute necessity, especially for portable devices whose lifetime is generally battery limited. However, it is unlikely to meet all the demands and is often infeasible to enhance all system parameters simultaneously. This is the point where circuit and system designers seek a compromise solution. In the last decade, various compromising methods have been presented to cope with the resource limitations on the one hand, and sustain high-performance operations on the other hand. Approximate computing is a design paradigm that exploits the intrinsic error-resilient trait of an application to achieve a trade-off between performance and computation quality [2]. It can be applied to diverse levels, from circuits upwards to compilers, algorithms, and even program languages [1, 3]. IBM and ARM have already considered approximate computing for their on-chip artificial intelligence accelerator and low-power approximate processor, respectively [4]. The utilization of imprecise techniques has also been reported in GPUs and FPGA memorization [5, 6].

Humans perceive the outside world mainly through vision and hearing. However, many of the multimedia applications interpreted by the human senses do not require full-precision results because of our limited perceptual capabilities [7]. Therefore, several parameters of the circuits related to hardware-level approximation, which mainly targets arithmetic units [8], can be improved by acceptable accuracy degradation. Image and signal processing have been among fruitful areas for the employment of imprecise arithmetic circuits such as adders [9-12] and multipliers [12-21].

A multiplier is an arithmetic unit with a considerable impact on the performance of the entire system. It is widely utilized in digital signal processing, linear algebra, cryptosystems, and 3D graphic accelerators [22-24]. Column compression (CC) multipliers are composed of three phases: 1) Partial product generation; 2) Partial product reduction; and 3) Final addition. The long critical path of a multiplier has always been a severe challenge for VLSI system designers. Moreover, there are many partial products in the second phase of multiplication to be accumulated. The CC multiplication, where partial products are summed up in a parallel manner, is a viable solution to the demand for high-speed multipliers and is known to be time-optimal [25].

After partial product generation in the first phase, it takes a few stages to reduce them by means of compressors in the second phase. The simplest form of a compressor is full adder. Wallace [26] and Dadda [27] are the two renowned classical CC multipliers that repeatedly utilize half and full adders to the point where only one or two partial products remain in each column. Then, a ripple-carry adder computes the final result in the last multiplication phase. Wider compressors can also be exploited for fast partial product reduction [28]. 4:2 compressor is the most common amongst various high-order compressors. Several varieties of 4:2 inexact compressors have also been suggested [14-21, 29]. There are some wider inexact compressors in the literature as well [13, 30-33]. Finally, truncation is another technique for constructing efficient approximate multipliers [13, 14].

This paper presents a novel multicolumn 3,3:2 inexact compressor. Then, two ultra-fast approximate multipliers are realized by the presented inexact compressor and its smaller derivatives. The new multicolumn compressor, which takes partial products from two adjacent columns, provides a unique situation under which the final result of the multipliers is achieved in only two stages. To the best of our knowledge, none of the previous approximate multipliers can accumulate partial products in such a few stages [13-21, 32-37]. Afterward, the proposed multipliers are utilized in the image sharpening application, which is one of the fundamental operations to improve visual effects and highlight fine details in an image [38].

The rest of the paper is organized as follows: The new multicolumn 3,3:2 inexact compressor is introduced in Section 2. Then, the proposed approximate multipliers through systematic analysis of performance and assessment of error are suggested in Section 3. Hardware-level and application-level implementation results are brought in Section 4. Finally, Section 5 concludes the paper.

## II. Proposed Multicolumn Inexact Compressor

Compressors in many different shapes and types have widely been employed in multipliers for rapid partial product accumulation. They perform addition and compression at the same time. An M:N compressor ($N<M$) takes M equally weighted partial products and returns their summation in the form of an N-digit number. In addition, it might have some input/output carries. To avoid horizontal carry propagation, input carries must not delay the production of output ones. For this purpose, output carries are often considered independent of input carries, although it is not a concrete obligation.

A multicolumn compressor can take partial products with different weights from more than one column. The general block diagram of a multicolumn compressor is shown in Fig. 1, where $M_i$ ($k \le i \le k+L-1$) indicates the number of input partial products from column $2^i$, P is the number input/output carries, and L implies the number of columns from which partial products are taken (starting from column $2^k$). This general block diagram can model either an exact or an inexact compressor. Additionally, it can also represent a single-column compressor by considering $L=1$.

Figure 2 shows the proposed multicolumn 3,3:2 inexact compressor. It takes six partial products from two neighboring columns, $2^k$ ($a_1$, $a_2$, and $a_3$) and $2^{k+1}$ ($b_1$, $b_2$, and $b_3$). In comparison with the equivalent 6:2 exact compressor (Fig. 3) [39], the new one has fewer hardware components, shorter critical path, and fewer carry signals; however, at the cost of producing erroneous outputs. The truth table of the proposed 3,3:2 inexact compressor is given in Table 1. The seven input variables, including $C_{in}$, lead to $2^7$=128 distinct input combinations, and thus 128 rows for the corresponding truth table. Nevertheless, the number of rows is reduced by the sum of input variables, Σin, which is calculated by (1). Every possible permutation of $\sum_i a_i = a_1 + a_2 + a_3$, $\sum_i b_i = b_1 + b_2 + b_3$, and $C_{in}$ that results in the same value of Σin is also demonstrated in Table 1. For example, both $(\sum_i b_i = 3, \sum_i a_i = 3, C_{in} = 0)$ and $(\sum_i b_i = 3, \sum_i a_i = 2, C_{in} = 1)$ cause Σin to be 9. Moreover, the probability of each row, P(row), is included in Table 1 to consider the effect of the entire input combinations in future calculations.

For a compressor, if the proportion of $M = \sum_{i=0}^{L-1} M_i$ to N increases, higher compression rate usually at the price of more latency is achieved. For the concurrent evaluation of these factors, a figure of merit, $FOM_1$ (2), is suggested in [39]. Smaller values of $FOM_1$ indicate faster operation and higher bit compression. However, $FOM_1$ reflects nothing about the amount of error in an inexact compressor. In 48, out of 128, rows of Table 1, at least one of the outputs is wrong. The error distance (ED) in these rows, the difference between exact and inexact output values (3), is −2 or −4. Another figure of merit, $FOM_2$ (4), with the inclusion of delay, power, and normalized error distance (NED) has been presented in [14].

The NED of a compressor ($NED_C$), (5), is equal to its mean error distance ($MED_C$) divided by the maximum possible ED, which happens in a hypothetical situation where all of the input variables are '1' and all of the output signals are '0'. The $MED_C$ parameter itself is the accumulation of absolute EDs divided by the number of input patterns (6). In a truth table representation by the probability of its rows, like the one exhibited in Table 1, $MED_C$ equals the sum of absolute EDs multiplied by their probabilities (6). Finally, (5) and (6) are adapted in this paper in such a way that the NED and MED of a multicolumn compressor can be calculated as well.

$$\sum in = \left( \sum_i b_i \right) \times 2^1 + \left( \sum_i (a_i) + C_{in} \right) \times 2^0 \quad (1)$$

$$FOM_1 = \frac{Delay}{\log(M) - \log(N)} \quad (2)$$

$$ED = Output_{exact} - Output_{inexact} \quad (3)$$

$$FOM_2 = \frac{Delay \times Power}{1 - NED_C} \quad (4)$$

$$NED_C = \frac{MED_C}{\max(error)} = \frac{MED_C}{\sum_{i=0}^{L-1} \left( M_i \times 2^i \right) + P} \quad (5)$$

$$MED_C = \frac{\sum_{i=1}^{2^{\sum_{j=0}^{L-1}(M_j)+P}} \left| ED_i \right|}{2^{\sum_{j=0}^{L-1}(M_j)+P}} = \sum_i \left( \left| ED_i \right| \times P(row = i) \right) \quad (6)$$

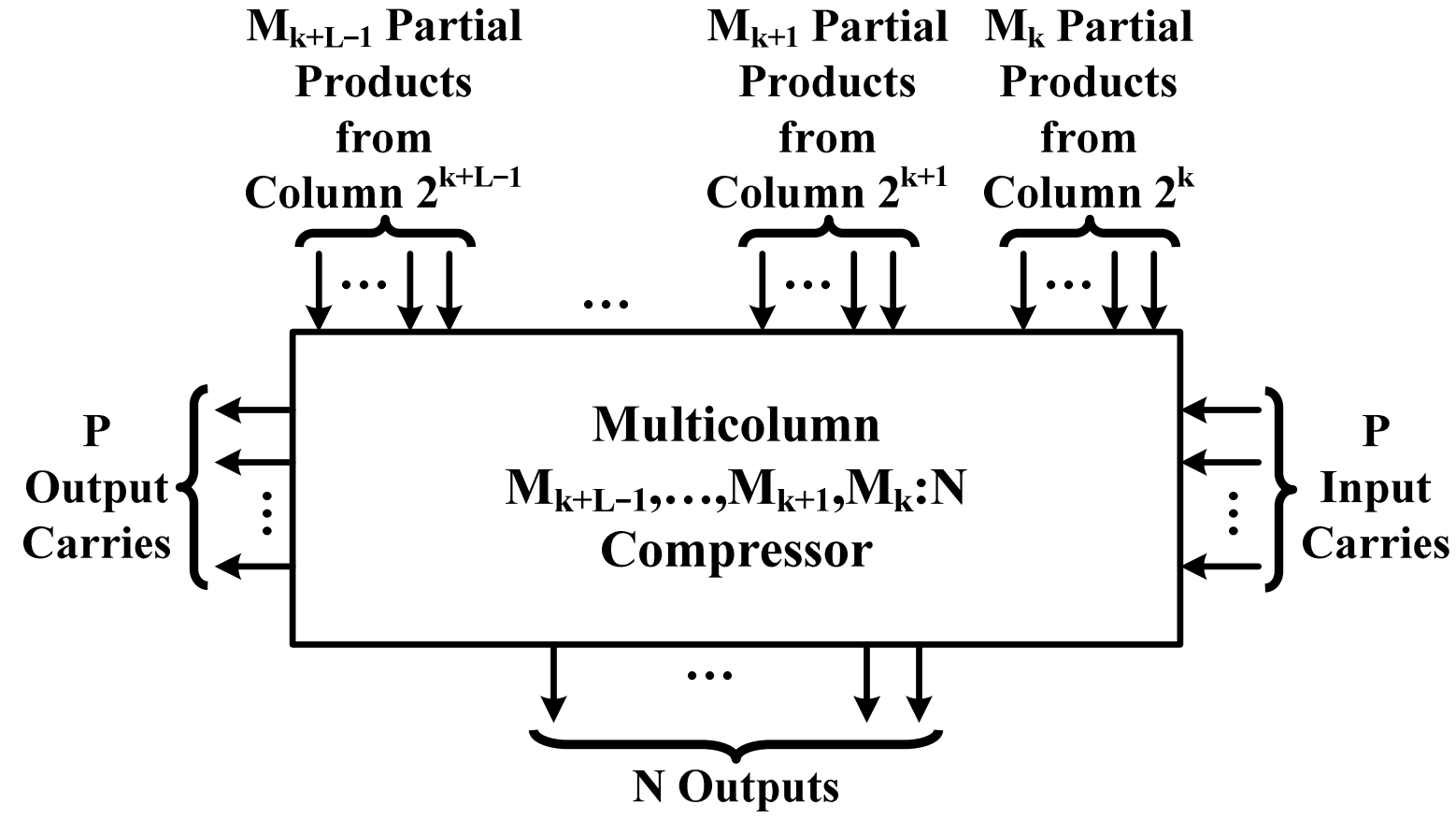

Fig. 1. The general block diagram of a multicolumn compressor.

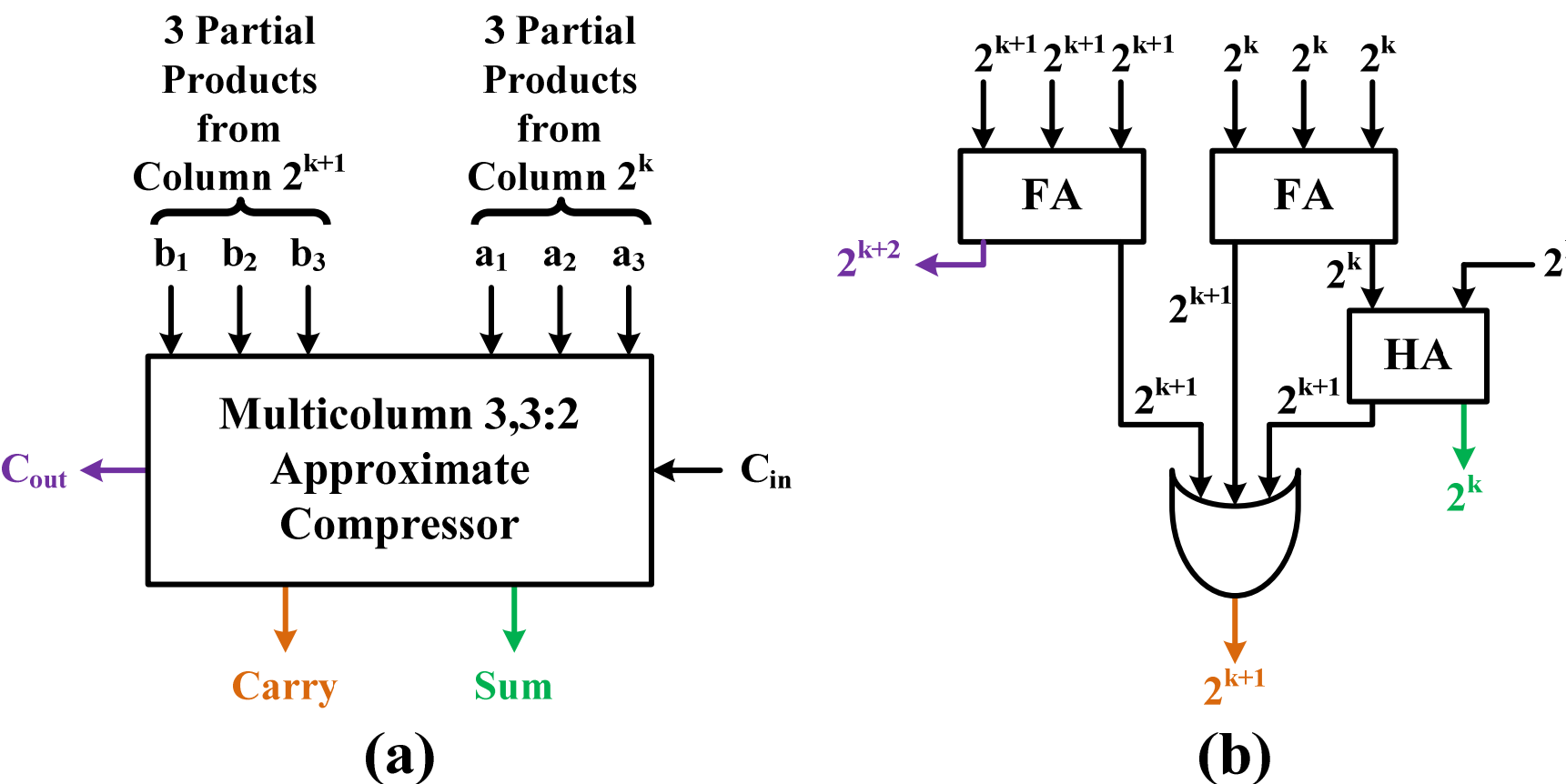

Fig. 2. The proposed multicolumn 3,3:2 inexact compressor, (a) Block diagram, (b) Inner structure.

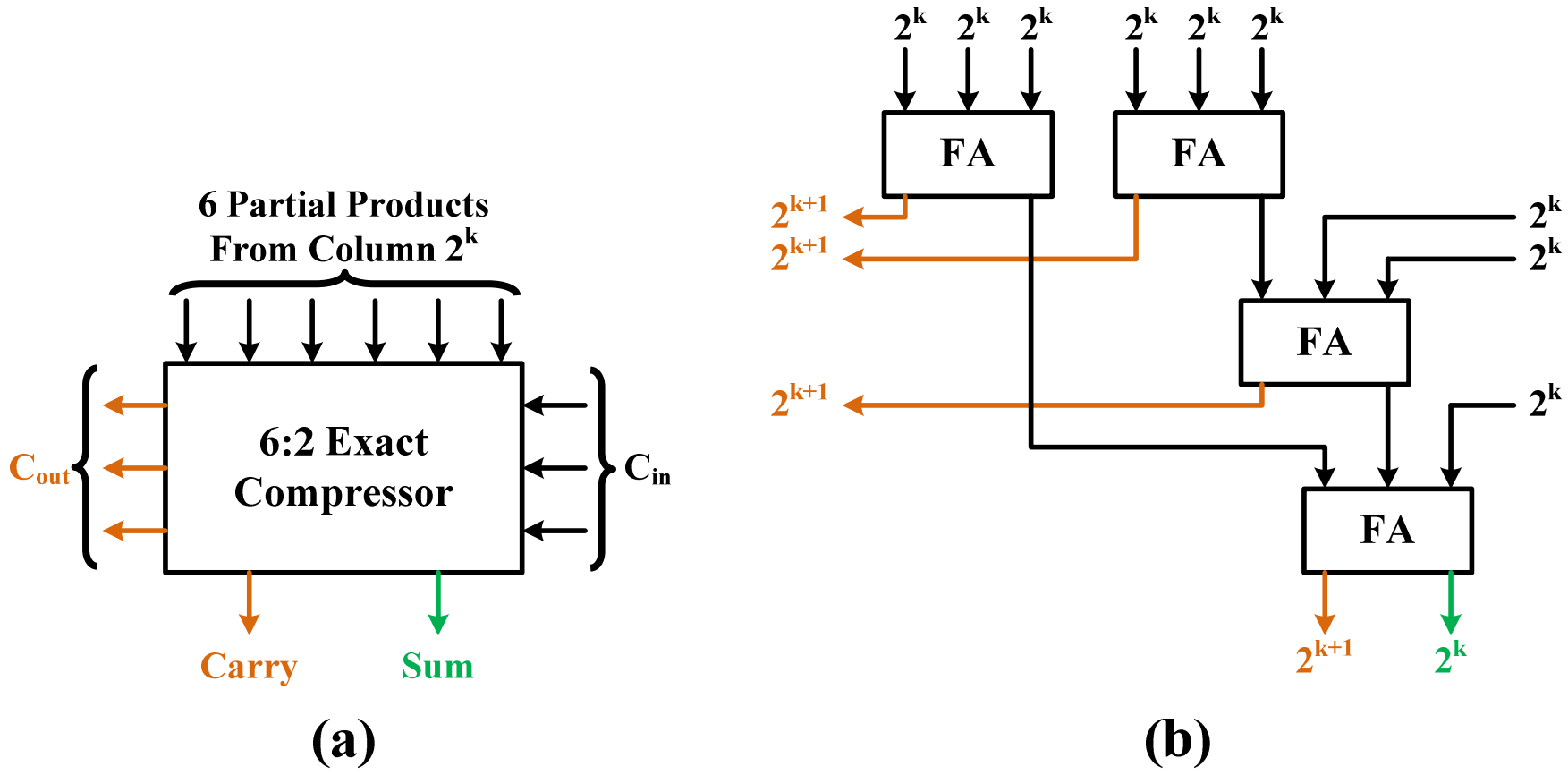

Fig. 3. The 6:2 exact compressor presented in [39], (a) Block diagram, (b) Inner structure.

Table 2 compares the proposed multicolumn compressor with some previously presented inexact designs in terms of NED, $FOM_1$, and $FOM_2$. The delay and power consumption parameters are obtained by Synopsys Design Complier and 45 nm technology node in the 1V power supply. The 4:2 inexact compressor is the most prevalent one in the literature. The ones in [14, 16-20] remove the carry signals. One of the input partial products is also truncated in [14]. As a result of this truncation, the compressor in [14] has a higher NED than the other equivalents. The highest

NED belongs to [13], a 5:2 inexact compressor, and the lowest NED belongs to [19, 20]. The NED of the proposed design stands in the middle. Furthermore, the values of $FOM_1$ show that the proposed compressor operates faster than [17, 19, 20], but not the other competitors, which means its performance in terms of $FOM_2$ is not satisfactory. However, we will see how the tables are turned in the next section, when the proposed design is utilized to construct a multiplier.

Although $FOM_1$ and $FOM_2$ give a preliminary estimate of how suitable an individual compressor is, they do not provide further information about its exact efficiency inside a multiplier. The given multicolumn compressor has the highest value of $FOM_2$ because it is composed of three accurate elements, two full adders and a half adder, which consume more power than the imprecise components. Nonetheless, the new design reduces the partial products of two adjacent columns concurrently. This feature brings some advantages, which are not comprehensible by NED, $FOM_1$, or $FOM_2$, for the new design over a 4:2 inexact compressor. Some examples are provided in Figs. 4 to 6 to demonstrate how the proposed compressor excels at either reducing partial products by fewer compressors or generating fewer partial products at the next stage.

Table 1. Truth Table of the Proposed Multicolumn 3,3:2 Inexact Compressor

| Σin | $\sum_i b_i$ | $\sum_i a_i$ | $C_{in}$ | $C_{out}$ | Carry | Sum | ED | P(row) |
|---|---|---|---|---|---|---|---|---|
| 0 | 0 | 0 | 0 | 0 ✓ | 0 ✓ | 0 ✓ | 0 | 1/128 |
| 1 | 0 | 0 | 1 | 0 ✓ | 0 ✓ | 1 ✓ | 0 | 1/128 |
| | 0 | 1 | 0 | 0 ✓ | 0 ✓ | 1 ✓ | 0 | 3/128 |
| 2 | 0 | 1 | 1 | 0 ✓ | 1 ✓ | 0 ✓ | 0 | 3/128 |
| | 0 | 2 | 0 | 0 ✓ | 1 ✓ | 0 ✓ | 0 | 3/128 |
| | 1 | 0 | 0 | 0 ✓ | 1 ✓ | 0 ✓ | 0 | 3/128 |
| 3 | 0 | 2 | 1 | 0 ✓ | 1 ✓ | 1 ✓ | 0 | 3/128 |
| | 0 | 3 | 0 | 0 ✓ | 1 ✓ | 1 ✓ | 0 | 1/128 |
| | 1 | 0 | 1 | 0 ✓ | 1 ✓ | 1 ✓ | 0 | 3/128 |
| | 1 | 1 | 0 | 0 ✓ | 1 ✓ | 1 ✓ | 0 | 9/128 |
| 4 | 0 | 3 | 1 | 0 × | 1 × | 0 ✓ | −2 | 1/128 |
| | 1 | 1 | 1 | 0 × | 1 × | 0 ✓ | −2 | 9/128 |
| | 1 | 2 | 0 | 0 × | 1 × | 0 ✓ | −2 | 9/128 |
| | 2 | 0 | 0 | 1 ✓ | 0 ✓ | 0 ✓ | 0 | 3/128 |
| 5 | 1 | 2 | 1 | 0 × | 1 × | 1 ✓ | −2 | 9/128 |
| | 1 | 3 | 0 | 0 × | 1 × | 1 ✓ | −2 | 3/128 |
| | 2 | 0 | 1 | 1 ✓ | 0 ✓ | 1 ✓ | 0 | 3/128 |
| | 2 | 1 | 0 | 1 ✓ | 0 ✓ | 1 ✓ | 0 | 9/128 |
| 6 | 1 | 3 | 1 | 0 × | 1 ✓ | 0 ✓ | −4 | 3/128 |
| | 2 | 1 | 1 | 1 ✓ | 1 ✓ | 0 ✓ | 0 | 9/128 |
| | 2 | 2 | 0 | 1 ✓ | 1 ✓ | 0 ✓ | 0 | 9/128 |
| | 3 | 0 | 0 | 1 ✓ | 1 ✓ | 0 ✓ | 0 | 1/128 |
| 7 | 2 | 2 | 1 | 1 ✓ | 1 ✓ | 1 ✓ | 0 | 9/128 |
| | 2 | 3 | 0 | 1 ✓ | 1 ✓ | 1 ✓ | 0 | 3/128 |
| | 3 | 0 | 1 | 1 ✓ | 1 ✓ | 1 ✓ | 0 | 1/128 |
| | 3 | 1 | 0 | 1 ✓ | 1 ✓ | 1 ✓ | 0 | 3/128 |
| 8 | 2 | 3 | 1 | 1 × | 1 × | 0 ✓ | −2 | 3/128 |
| | 3 | 1 | 1 | 1 × | 1 × | 0 ✓ | −2 | 3/128 |
| | 3 | 2 | 0 | 1 × | 1 × | 0 ✓ | −2 | 3/128 |
| 9 | 3 | 2 | 1 | 1 × | 1 × | 1 ✓ | −2 | 3/128 |
| | 3 | 3 | 0 | 1 × | 1 × | 1 ✓ | −2 | 1/128 |
| 10 | 3 | 3 | 1 | 1 × | 1 ✓ | 0 ✓ | −4 | 1/128 |

When there are four partial products within each column, three proposed compressors are adequate to get the same result as the six 4:2 compressors do. This scenario is exemplified in Fig. 4. In the following example (Fig. 5), where there are five partial products in each column, the combined usage of the presented compressor and one of its derivatives, called 2,2:2, results in eliminating an entire row of partial products at the next stage. The multicolumn 2,2:2 inexact compressor (Fig. 5(c)) is simply achieved by replacing the full adders in Fig. 2(b) with half adders. Not only does the replacement simplify the inner structure, but it also leads to the production of more precise results. The sum of input variables never becomes 8, 9, and 10 in the 2,2:2 compressor. $\Sigma in=8$, $\Sigma in=9$, and $\Sigma in=10$ are the situations in Table 1 where the results are always incorrect. As a result of their elimination, $NED_C$ decreases from 0.08125 for 3,3:2 to 0.07143 for 2,2:2. In the last example (Fig. 6), where there are six partial products inside each column, the proposed compressors generate two fewer rows of partial products at the next stage.

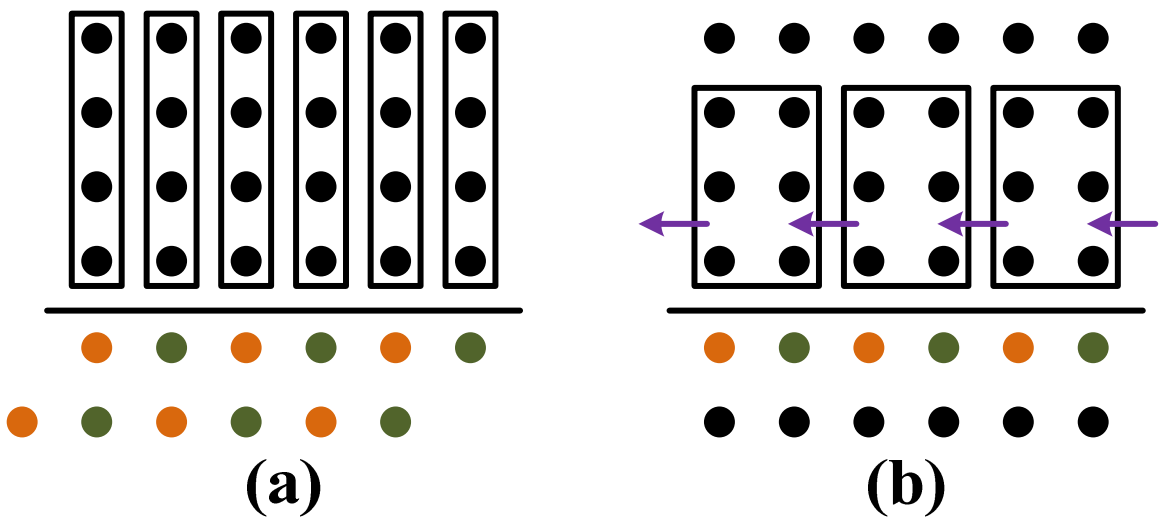

Fig. 4. Accumulation of six columns with four partial products each, (a) By 4:2 inexact compressors, (b) By the proposed multicolumn 3,3:2 inexact compressors.

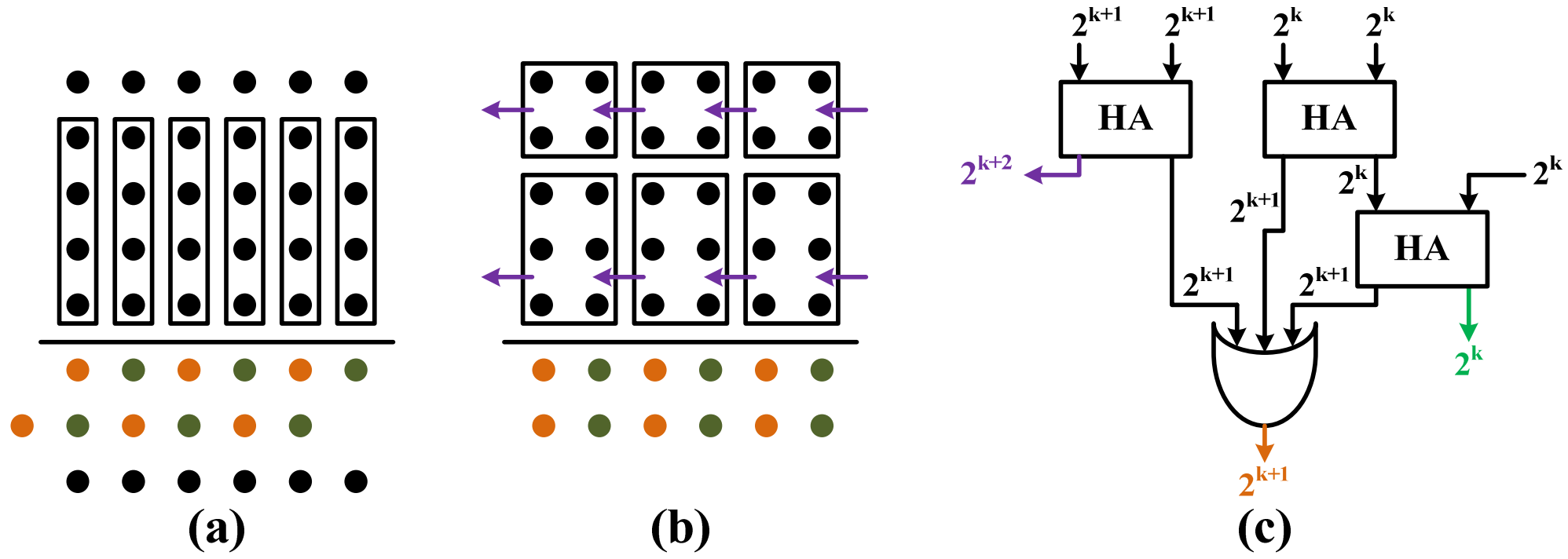

Fig. 5. Accumulation of six columns with five partial products each, (a) By 4:2 inexact compressors, (b) By the proposed multicolumn 3,3:2 and 2,2:2 inexact compressors, (c) Inner structure of the 2,2:2 inexact compressor.

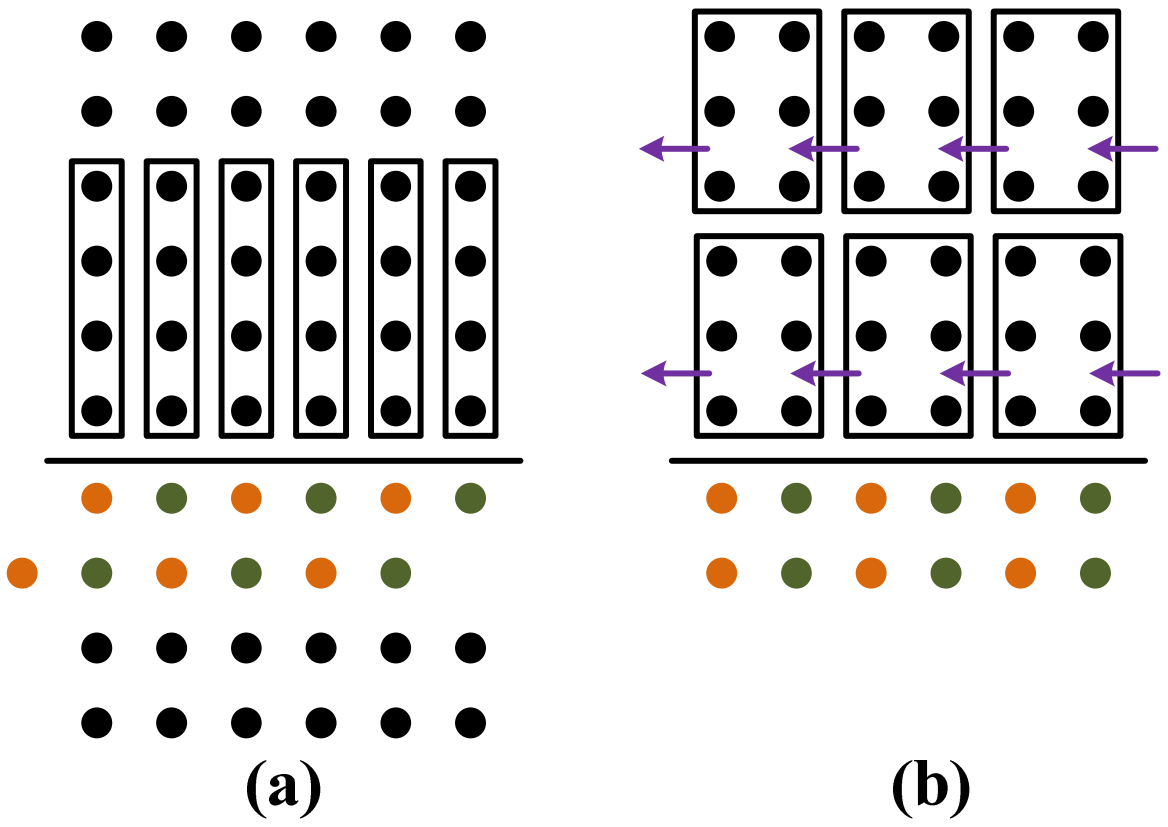

Fig. 6. Accumulation of six columns with six partial products each, (a) By 4:2 inexact compressors, (b) By the proposed multicolumn 3,3:2 inexact compressors.

Table 2. Comparison of Inexact Compressors

| Compressor | Inner Structure | NED | $FOM_1$ | $FOM_2$ |
|---|---|---|---|---|
| (3,3:2) New | | 0.08125 | 0.482 | 4.743 |
| 5:2 [13] | | 0.13125 | 0.251 | 0.258 |
| 4:2 [14] | | 0.125 | 0.199 | 0.100 |
| 4:2 [15] | | 0.075 | 0.365 | 0.444 |
| 4:2 [16] | | 0.078125 | 0.432 | 0.848 |
| 4:2 [17] | | 0.078125 | 0.698 | 1.018 |
| 4:2 [18] | | 0.078125 | 0.432 | 0.848 |
| 4:2 [19] | | 0.03125 | 0.698 | 2.059 |
| 4:2 [20] | | 0.03125 | 0.565 | 1.761 |
| 4:2 [21] | | 0.1 | 0.199 | 0.097 |

## III. Proposed Approximate Multipliers

Figure 7 shows the initial design of the proposed 8×8 approximate multiplier, whose structure merely relies on the proposed multicolumn 3,3:2 inexact compressor and its first derivative, i.e. 2,2:2. It receives two 8-bit unsigned numbers, $A_{7\text{-}0}$ and $B_{7\text{-}0}$, and returns its 16-bit product, $F_{15\text{-}0}$. There is not any precise component in the initial design. The unique feature of this multiplier is that it completes the entire partial product reduction (the second and third multiplication phases) in only two stages. The second and third multiplication phases are merged in the proposed design. Despite its apparent efficiency, it is still far from being entirely suitable because of two significant reasons:

1. The presented multiplier is not in its simplest form yet. Some of the 3,3:2 and 2,2:2 compressors are loaded with zeros, which certainly do not affect the addition result. Therefore, we can further simplify the initial design by defining other derivatives of the proposed multicolumn 3,3:2 compressor.
2. The error distance of the final result, $F_{15\text{-}0}$, becomes extremely large if the imprecise components are inserted into the most significant columns. For instance, the most significant bit (MSB) of the final output, $F_{15}$, is currently always equal to '0'. It varies the output by about $2^{15}$=32768 whenever F is supposed to be greater than this number.

Precise components such as half adder, full adder, and 4:2 compressor are slower, larger, and less energy-efficient than their imprecise counterparts. However, they produce correct outputs. The question is how many columns in higher bit positions should be occupied by the precise components to make a trade-off between accuracy and hardware efficiency metrics such as delay, power consumption, and area. Instead of a heuristic approach, a systematic investigation is carried out in this paper to find the answer.

We start increasing the accuracy of our initial design (Fig. 7) by adding some precise components to the most significant columns of Stage #1. Figure 8 shows seven approximate multipliers, each of which has one precise component more than the previous one in the chain of successive precise components at Stage #1. After inserting a specific number of precise components into the first stage, the remaining partial products are fed to the proposed multicolumn 3,3:2 inexact compressor or its various derivatives, illustrated in Appendix I (Table 6).

At the first stage, the fewest possible compressors are utilized to generate no more than three partial products at Stage #2. In addition, the derived inexact compressors are leveraged to simplify the structure of the multipliers to a greater degree. Regarding $FOM_1$, not all of the derivatives are better than the initial 3,3:2 compressor because $FOM_1$ depends on only compression rate and delay. However, when it comes to $FOM_2$, all of them show better performance than the primary version. At Stage #2, the proposed compressors in lower bit positions and a ripple-carry adder in higher bit positions compute the final result.

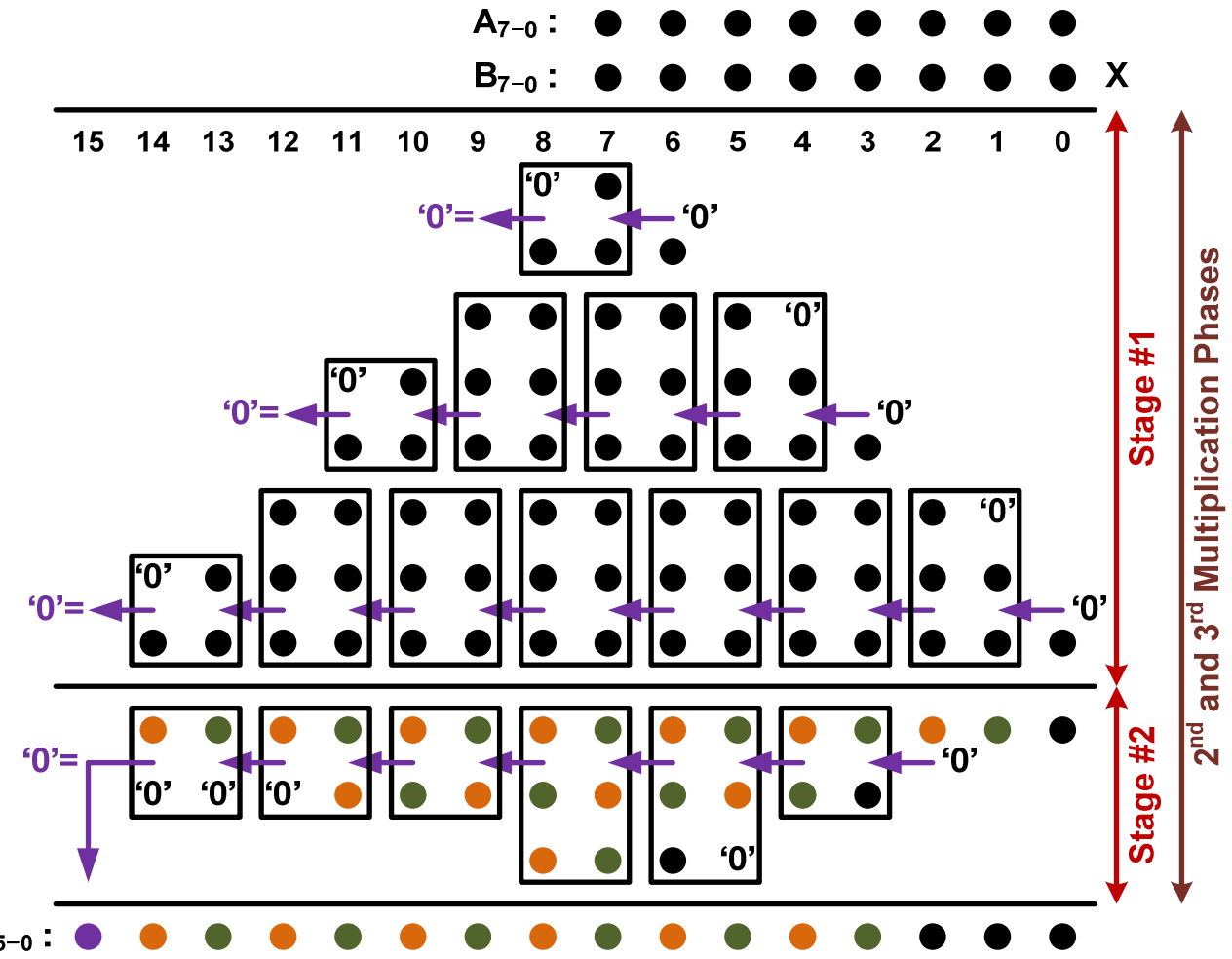

Fig. 7. The proposed approximate multiplier (Initial design), which completes the second and third multiplication phases in only two stages.

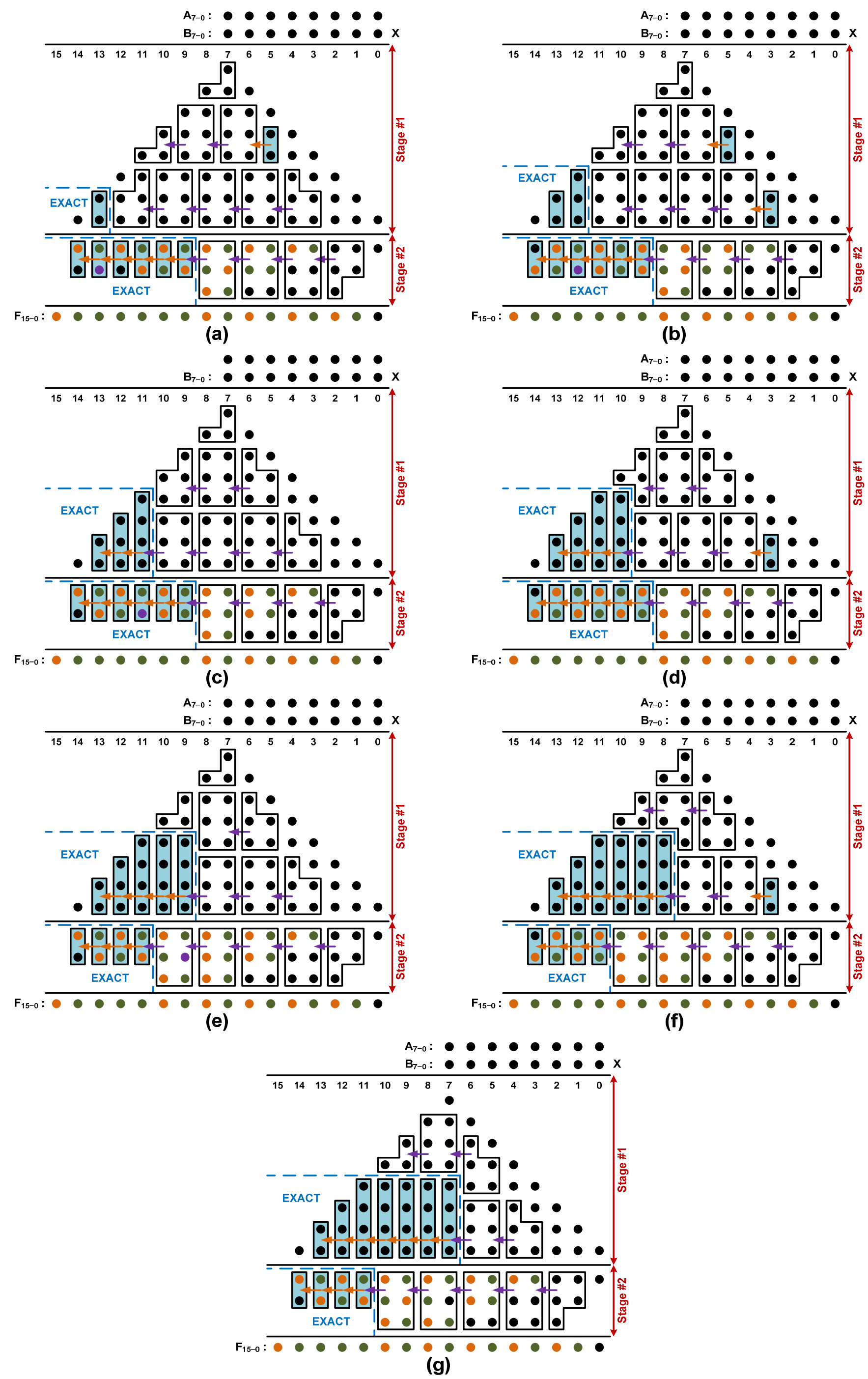

Fig. 8. 8×8 approximate multipliers by the proposed multicolumn 3,3:2 inexact compressor and its derivatives when there are a number of precise components at Stage #1, (a) One, (b) Two, (c) Three, (d) Four (Design #1), (e) Five, (f) Six, (g) Seven.

At the first stage of the first approximate multiplier (Fig. 8(a)), a half adder, as a precise component, is inserted into column 13. In the second approximate multiplier (Fig. 8(b)), a full adder and a half adder are put in columns 12 and 13, respectively. In the third to seventh multipliers (Figs. 8(c) to 8(g)), exact 4:2 compressors are utilized to create a chain of precise components at Stage #1. Besides, the previous full and half adders in columns 12 and 13 are replaced with a 4:2 compressor (with three input partial products) and a full adder, respectively, to avoid sending the output carry of the 4:2 compressor in column 11 to the next stage.

The chain of precise components at Stage #1 cannot exceed seven without violating the production of maximum of three partial products at Stage #2. Therefore, Fig. 8(g) is the multiplier with the longest chain of precise components. Additionally, the size of the ripple-carry adder at Stage #2 depends on the arrangement of Stage #1. The last three multipliers, Figs. 8(e) to 8(g), have shorter ripple-carry adders than the first four ones, Figs. 8(a) to 8(d). Moreover, a few half adders are rarely utilized in the least significant columns to adhere to the policy of no more than three partial products at Stage #2. Finally, all of the precise components in Fig. 8 are filled with light blue to distinguish them from inexact compressors.

The MED of an n×n multiplier, $MED_M$, is the accumulation of absolute EDs divided by the number of input patterns, i.e. $2^n \times 2^n$ (7). Furthermore, $NED_M$ is equal to $MED_M$ divided by the maximum possible ED, i.e. $(2^n-1)\times(2^n-1)$ for an n×n multiplier (8). The product of power, delay, area, and MED (PDAEP) is considered as a measure of appropriateness for the multipliers (9). All of these parameters are supposed to be reduced, and thus, lower values of PDAEP are certainly preferred.

Figure 9 shows the values of PDAEP for the approximate multipliers depicted in Fig. 8. The one with four successive precise components at Stage #1 (Fig. 8(d)) has the lowest PDAEP value. Therefore, we consider it to be our first proposed design (Design #1). The last three multipliers with five, six, and seven successive precise components have slightly higher PDAEP values because of their shorter ripple-carry adders at Stage #2. As the size of the ripple-carry adder shrinks in the last three multipliers, MED increases even though there are longer series of precise components at their first stage. Therefore, there is a sudden increase in MED for the fifth multiplier. The multiplier accuracy relies on how many columns in higher bit positions are occupied by precise components at all stages.

Truncation of partial products is a beneficial technique for constructing an energy-efficient approximate multiplier. It eliminates several compressors from the least significant columns and leads to fewer AND gates during partial product generation. The effects of the truncation strategy on the proposed multiplier are again systematically studied by a gradual increase in the number of truncated columns. Figure 10 shows the first proposed approximate multiplier, Design #1, with different numbers of truncated columns. The MED and power-delay-area product (PDAP) (10) values of the approximate multipliers are simultaneously plotted in Fig. 11. As the number of truncated columns increases, MED increases while PDAP decreases. Their intersection, which stands in between five and six, is the point with the maximum efficiency. Multipliers with five (Fig. 10(e)) or six (Fig. 10(f)) truncated columns show approximately similar performances. The latter is considered to be our second proposed design (Design #2) for its more hardware efficiency.

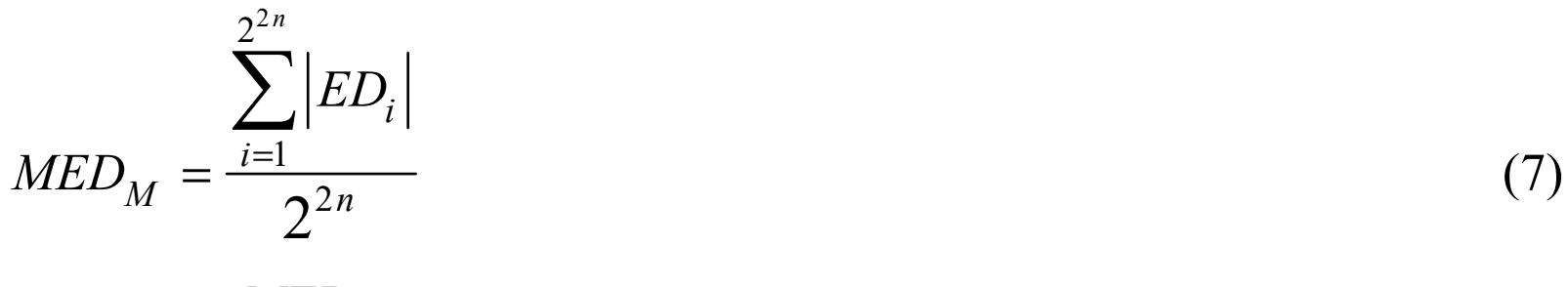

$$MED_M = \frac{\sum_{i=1}^{2^{2n}} |ED_i|}{2^{2n}} \quad (7)$$

$$NED_M = \frac{MED_M}{(2^n-1)^2} \quad (8)$$

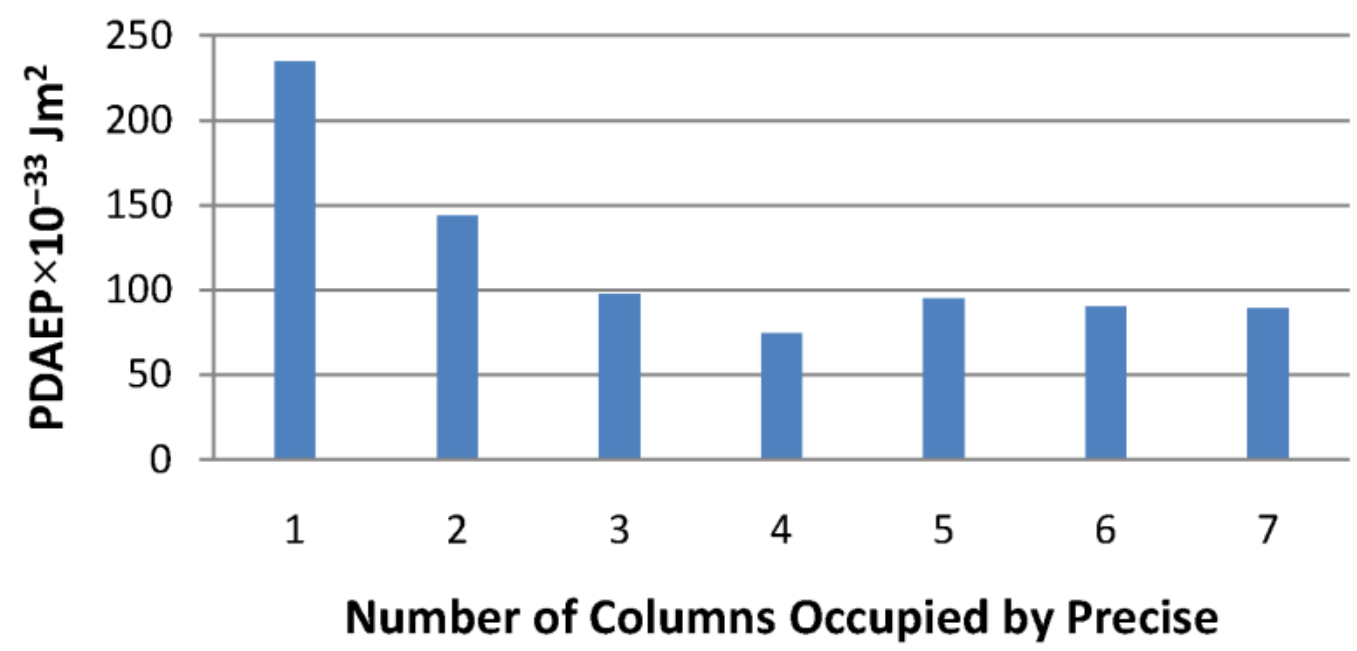

Fig. 9. PDAEP vs. the number of columns occupied by precise components for the illustrated multipliers in Fig. 8.

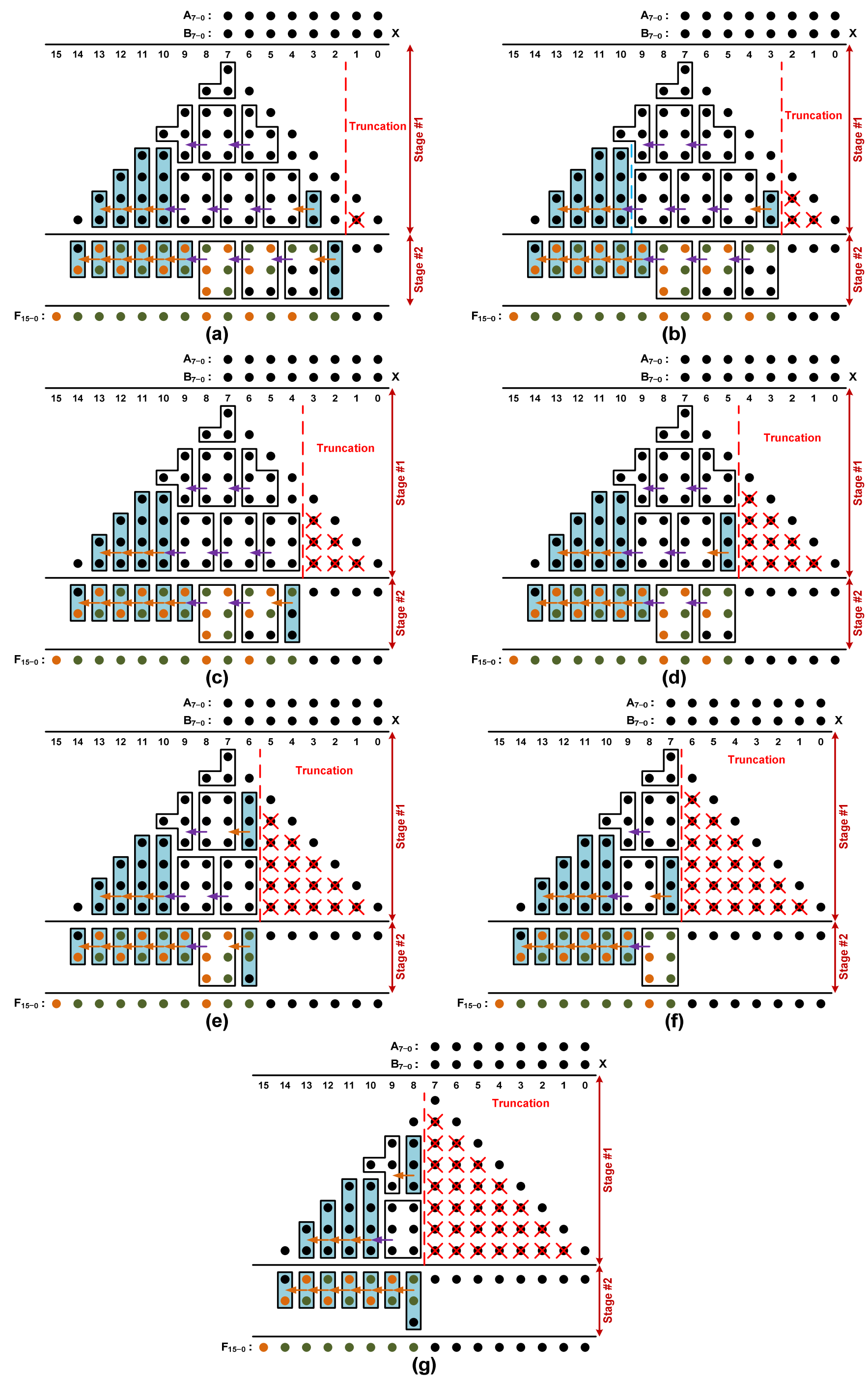

Fig. 10. The first proposed 8×8 approximate multiplier, Design #1, with different numbers of truncated columns, (a) One, (b) Two, (c) Three, (d) Four, (e) Five, (f) Six (Design #2), (g) Seven.

$$PDAEP = (P_{Dynamic} + P_{Leakage}) \times Delay \times Area \times MED \tag{9}$$

$$PDAP = (P_{Dynamic} + P_{Leakage}) \times Delay \times Area \tag{10}$$

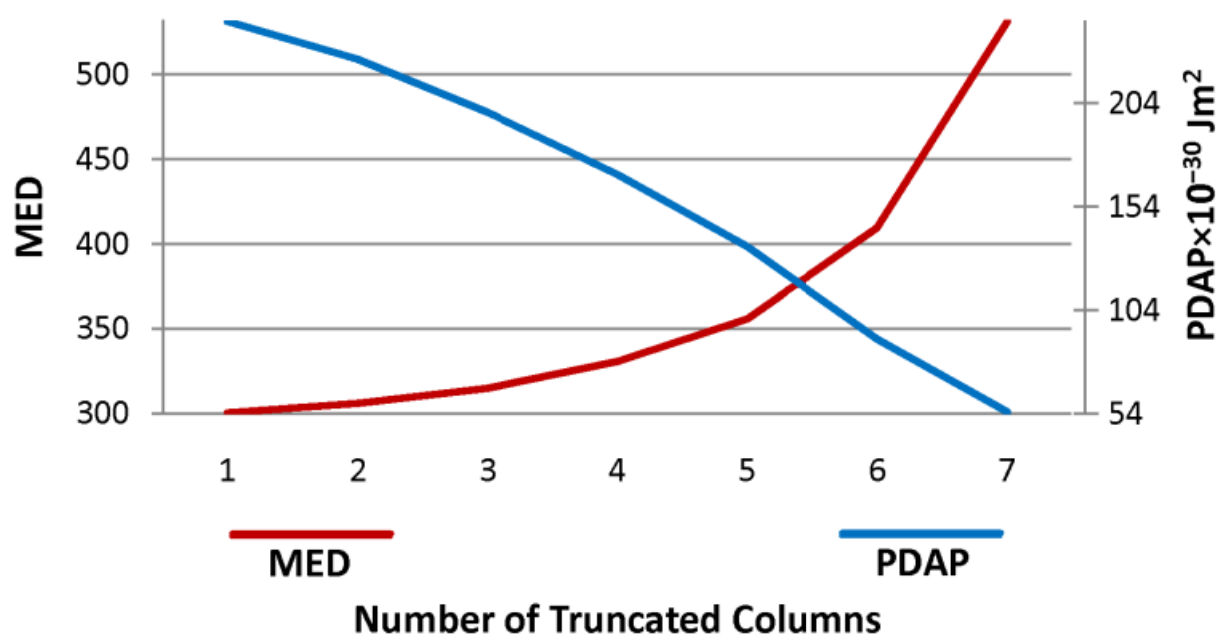

Fig. 11. MED and PDAP vs. the number of truncated columns for the illustrated multipliers in Fig. 10.

Although the initial results of the new multicolumn 3,3:2 inexact compressor were not desirable, the two proposed approximate multipliers, Design #1 and Design #2, seem to be very promising because:

1. It takes only two stages for the proposed designs to compute the final result. As far as we know, this unique feature is not available in any of the previous approximate multipliers.
2. The derived compressors, which substantially contribute to the process of partial product reduction in the proposed multipliers, are faster, smaller, and even more precise than the initial 3,3:2 version (Appendix I, Table 6).
3. The precise components in higher bit positions help to have reasonable accuracy.
4. The truncation of the least significant columns in the second design helps to save more energy and area.

## IV. Implementation Results

### *A. Hardware-Level Implementation Results*

Synthesizable structural VHDL code is used to describe the proposed approximate multipliers, Design #1 (Fig. 8(d)) and Design #2 (Fig. 10(f)). Then, they are implemented by Synopsys Design Complier and 45 nm technology node in 1V power supply. Implementation results are shown in Tables 3 and 4, where error rate (ER) means the frequency of error, and PDAEP (9), PDAP (10), and power-delay product (PDP) (11) are the comprising evaluating factors. While PDP and PDAP specify hardware-level parameters, PDAED gives an initial estimate for application-level uses by covering the element of error as well. Two previous accurate multipliers and some of the previous approximate multipliers [13-21] are also implemented under the same conditions.

Table 3 shows that the new designs are far better than the accurate multipliers. They completely outperform the accurate multiplier made by 6:2 exact compressors [40]. Additionally, the second proposed multiplier operates 36.5% faster, consumes 60.6% less power, and occupies 52.7% less area than Dadda's multiplier, one of the most high-performance accurate multipliers. Furthermore, the approximate multipliers are compared in Table 4, where the boldface text indicates the best results. In brief:

- Delay: Both of the proposed approximate multipliers outperform all of the previous designs in terms of speed, and they are the fastest approximate multipliers.
- Accuracy: In terms of MED and NED, the presented designs have higher accuracy than the previous multipliers, except [18, 19]. However, neither of the implementations in [18, 19] shows acceptable results regarding delay, power consumption, and area. They are the worst multipliers in terms of PDP and PDAP. Additionally, Design #1 has the lowest ER, again except [18, 19].
- Power: Although the suggested multipliers do not consume the least amount of power, their dynamic and leakage power dissipation are quite reasonable, compared to the other competitors.
- Area: Design #2 occupies the least area, nearly 9.7% smaller than [14], which is the second smallest multiplier.
- PDP: The previous multiplier in [21] has the lowest PDP factor. Design #2 is the runner-up.
- PDAP: When considering the area as well as delay and power, Design #2 is the outright winner and surpasses [21] by 10.9%.

- PDAEP: After [19], Design #2 has the highest performance in terms of PDAEP. It is worth mentioning that the one in [19] is almost an accurate multiplier. It is as slow as the Dadda's multiplier, and dissipates nearly the same amount of power and area as the Dadda's multiplier does.

$$PDP = (P_{Dynamic} + P_{Leakage}) \times Delay \tag{11}$$

Table 3. Comparison of the Proposed Approximate Multipliers with Accurate Multipliers

| Multiplier | Delay (ns) | Power (µW) | | Area (µm$^2$) | PDP (fJ) | PDAP $\times 10^{-30}$ Jm$^2$ |
|---|---|---|---|---|---|---|
| | | Dynamic | Leakage | | | |
| Dadda's Accurate Multiplier | 1.26 | 576.08 | 6.25 | 1040 | 733 | 763.07 |
| Accurate Multiplier by 6:2 Compressors [40] | 1.37 | 662.80 | 6.63 | 1112 | 917 | 1019.8 |
| Proposed Design #1 (Approximate) | 0.85 | 369.13 | 4.77 | 786 | 318 | 249.82 |
| Proposed Design #2 (Approximate) | 0.80 | 226.35 | 2.96 | 492 | 183 | 90.22 |

Table 4. Comparison of the Proposed Approximate Multipliers with Previous Approximate Multipliers

| Multiplier | MED* | NED* $\times 10^{-3}$ | ER (%) | Delay (ns) | Power (µW) | | Area (µm$^2$) | PDP (fJ) | PDAP $\times 10^{-30}$ Jm$^2$ | PDAEP $\times 10^{-33}$ Jm$^2$ |
|---|---|---|---|---|---|---|---|---|---|---|
| | | | | | Dynamic | Leakage | | | | |
| Proposed Design #1 | 297.9 | 4.58 | 66.9 | 0.85 | 369.13 | 4.77 | 786 | 318 | 249.82 | 74.43 |
| Proposed Design #2 | 409.7 | 6.30 | 94.5 | **0.80** | 226.35 | 2.96 | **492** | 183 | **90.22** | 36.96 |
| [13] | 1190 | 18.3 | 91.4 | 1.01 | 194.62 | 3.18 | 570 | 200 | 113.82 | 135.45 |
| [14] | 455.2 | 7.00 | 99.8 | 0.90 | 206.38 | **2.95** | 545 | 188 | 102.65 | 46.72 |
| [15] | 3480 | 53.5 | 99.8 | 1.15 | 267.90 | 4.06 | 758 | 313 | 237.19 | 825.46 |
| [16] | 1157 | 17.8 | 85.4 | 1.11 | 235.01 | 4.00 | 636 | 265 | 168.71 | 195.27 |
| [17] | 944.7 | 14.5 | 97.5 | 0.94 | 253.05 | 3.32 | 696 | 241 | 167.60 | 158.33 |
| [18] | 87.1 | 1.3 | 66.5 | 1.14 | 449.69 | 5.64 | 961 | 519 | 498.65 | 43.45 |
| [19] | **0.17** | **$2.61\times10^{-3}$** | **47.0** | 1.26 | 530.41 | 5.73 | 978 | 676 | 660.36 | **0.112** |
| [20] | 502.9 | 7.73 | 79.9 | 1.20 | 430.19 | 5.19 | 913 | 522 | 477.14 | 239.95 |
| [21] | 3291 | 50.6 | 99.1 | 1.11 | **149.61** | 3.04 | 598 | **169** | 101.31 | 333.39 |

* MED and/or NED of the previous approximate multipliers have been extracted from the original papers.

### *B. Application-Level Implementation Results*

In this section, an image-sharpening algorithm by utilizing the proposed approximate multipliers is applied to the samples publicly available in the Local Image Sharpness Database [41, 42]. The collection contains six colored images with 384×284 pixels in size. The sharpening kernel is a sliding window that is multiplied by the whole image pixels. For the multiplication part, the system multiplication function, as an accurate multiplier, is called. Then, the same procedure is repeated by replacing the system multiplier with one of the approximate multipliers coded in C++. The I/O image manipulation is done by Numpy, Scikit-image, and Pillow libraries in the Python 3.6 platform. The experiments are done on a Macbook Pro, using a 2.6 GHz 6-Core Intel Core i7 and 16 GB of RAM.

An image is a matrix of pixels, where each pixel is an 8-bit unsigned integer. The utilized image-sharpening algorithm is very similar to those in [43, 44], where the sharpening matrix S of a given image I is calculated by (12), where B is a Gaussian blur estimation of the original image [45]. Such a blurred version can be obtained by applying a kernel matrix, G (13), all over the pixels of a given image. In this case, B is described by (14).

$$S = I + 1.5 \times (I - B) \tag{12}$$

$$G = \begin{bmatrix} 1 & 4 & 7 & 4 & 1 \\ 4 & 16 & 26 & 16 & 4 \\ 7 & 26 & 41 & 26 & 7 \\ 4 & 16 & 26 & 16 & 4 \\ 1 & 4 & 7 & 4 & 1 \end{bmatrix} \tag{13}$$

$$B(x,y) = \frac{1}{273}\sum_{i=-2}^{2}\sum_{j=-2}^{2} G(i+3, j+3) I(x-i, y-j) \tag{14}$$

For image quality assessment, the peak signal-to-noise ratio (PSNR) and structural similarity index measure (SSIM) evaluation metrics are taken into account. In our experiments, the accurately sharpened image is compared to the ones generated by the approximate multipliers. PSNR, which is calculated by (15), is the ratio of the maximum possible value of a signal ($MAX_f$) to its distortion rate, which is measured by the root of the mean square error (MSE) of the pairs of the corresponding pixels within images $I_1$ and $I_2$ (16).
Furthermore, SSIM is a perceptual metric to quantify the similarity of two given images. While PSNR only focuses on the differences between corresponding pixels, SSIM is based on the visual structures inside an image. SSIM, which is calculated by (17), is the product of three key features of an image: 1) Luminance, denoted by l; 2) Contrast, denoted by c; and 3) Structure, denoted by s.
By setting α, β and γ to 1 in (17), and defining the luminance, contrast, and structure as the functions of mean and variance of image pixels, we reach the final equation of SSIM, (18), where μs are the average of pixels, σs are the variances, and $C_1$ and $C_2$ are two variables to stabilize a very small or zero denominator. SSIM varies between zero and one, indicating the lowest and highest similarities, respectively.

$$PSNR = 20\log_{10}\left(\frac{MAX_f}{\sqrt{MSE}}\right) \tag{15}$$

$$MSE = \frac{1}{n\times m}\sum_{i=1}^{n}\sum_{j=1}^{m}\left(I_1(i,j) - I_2(i,j)\right)^2 \tag{16}$$

$$SSIM(x,y) = \left[l(x,y)\right]^{\alpha} . \left[c(x,y)\right]^{\beta} . \left[s(x,y)\right]^{\gamma} \tag{17}$$

$$SSIM(x,y) = \frac{\left(2\mu_x\mu_y + C_1\right)\left(2\sigma_{xy} + C_2\right)}{\left(\mu_x{}^2 + \mu_y{}^2 + C_1\right)\left(\sigma_x{}^2 + \sigma_y{}^2 + C_2\right)} \tag{18}$$

The sharpened images for a blurry photo of a squirrel are demonstrated in Fig. 12. The output images for the rest of the photos in the Local Image Sharpness Database are also illustrated in Appendix II (Figs. 14 to 18). They enable us to judge the outcomes of the approximate multipliers visually. In addition, the numerical results of the image-sharpening algorithm for all the approximate multipliers in terms of PSNR and SSIM are shown in Table 5. Each row shows the similarity between the generated images by a specific approximate multiplier and the corresponding one produced accurately. The PSNR and SSIM values presented in Table 5 are the mean of evaluations of all six samples (the entire photo collection).
As it is clear in Table 5, the previous multiplier in [19] achieves the best results since it is almost an accurate multiplier. Despite its superiority in the application-level evaluation, its hardware-level performance is not satisfactory at all. We also observe that the multipliers in [14, 15, 20] fail to produce satisfactory results, and the functionality of these approximate multipliers is far below expectations. Therefore, further analysis is carried out to understand the reason behind such poor performances. We bring in a theory that the multipliers' resulting dark images are because they produce higher error values when small numbers are multiplied. The idea comes from the fact that the kernel, which is used in the image-sharpening algorithm, consists of small numbers.
To verify our hypothesis, for each approximate multiplier, a 256-by-256 matrix, where the cell (i, j) is filled by the absolute ED of the product of i and j, is generated. Then, the matrix is mapped onto a heatmap chart for data visualization. All of the heatmap charts are depicted in Fig. 13, where darker colors imply greater absolute errors. As visible in the heatmap charts, for those three failing multipliers [14, 15, 20], larger absolute errors occur on either the top or the left borders, where smaller numbers are multiplied. We can also observe that the same borders/areas in the multipliers with a desirable application-level performance, including the proposed ones, are almost light.

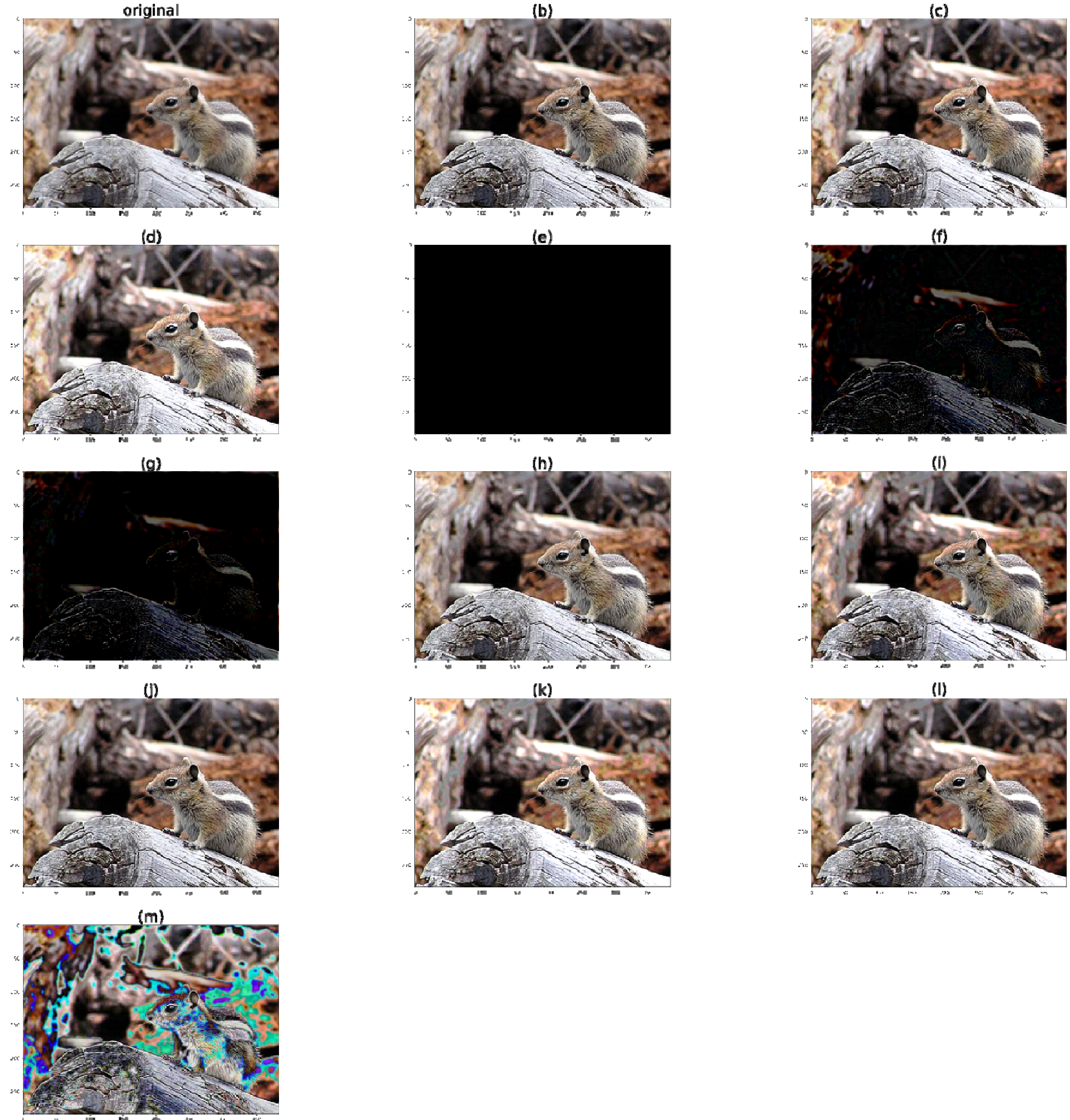

Fig. 12. Image of a squirrel, (a) Original, and sharpened by (b) System multiplier, (c) Proposed (Design #1), (d) Proposed (Design #2), (e) [15], (f) [20], (g) [14], (h) [18], (i) [17], (j) [19], (k) [16], (l) [13], (m) [21].

Table 5. The Average SSIM and PSNR Values

| Approximate Multiplier | SSIM | PSNR |
|---|---|---|
| Proposed Design #1 | 0.946871 | 28.29 |
| Proposed Design #2 | 0.892929 | 22.47 |
| [13] | 0.943720 | 31.86 |
| [14] | 0.039266 | 7.38 |
| [15] | $1\times10^{-6}$ | 6.69 |
| [16] | 0.910600 | 24.96 |
| [17] | 0.918917 | 30.51 |
| [18] | 0.984473 | 35.95 |
| [19] | 0.999428 | 60.9 |
| [20] | 0.159800 | 7.84 |
| [21] | 0.487409 | 10.28 |

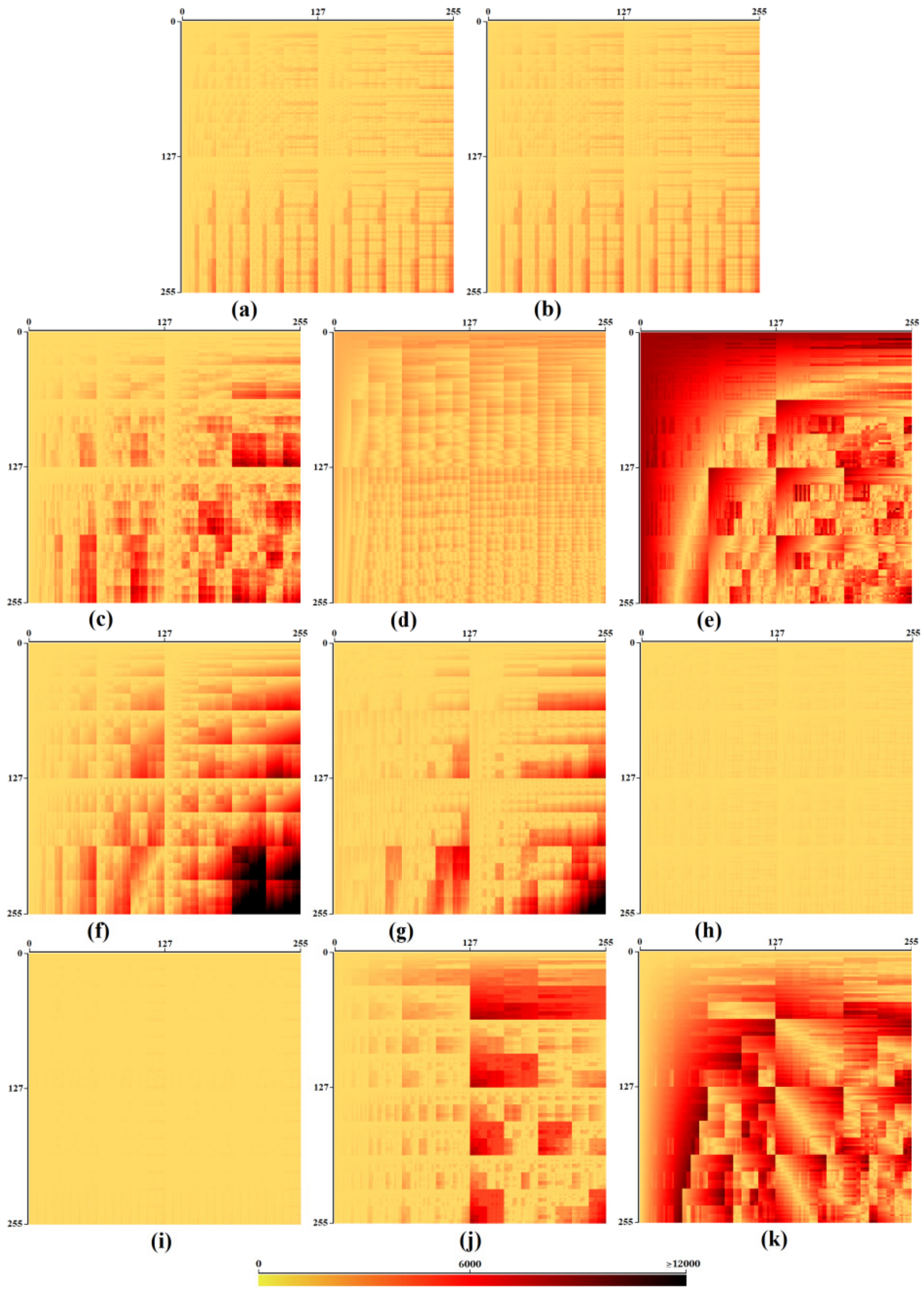

Fig. 13. Heatmap charts of the approximate multipliers, (a) Proposed (Design #1), (b) Proposed (Design #2), (c) [13], (d) [14], (e) [15], (f) [16], (g) [17], (h) [18], (i) [19], (j) [20], (k) [21].

## V. Conclusion and Future Works

Two new approximate multipliers based on a new multicolumn inexact compressor and its various derivatives have been presented. Implementation results are auspicious. The proposed multipliers are not only ultra-fast, but also quite efficient from other aspects. They produce desired outcomes, both visually and numerically, in the image sharpening application as well. We have reached three main conclusions based on our observations in this paper:

1. It is quite unfair to assess a compressor only by its individual performance. Despite initial unsatisfactory results (e.g. by $FOM_1$ and $FOM_2$), the final outcome might ironically be gratifying as the compressor gets involved in the structure of a multiplier. The proposed multicolumn inexact compressor is an excellent example of this case.
2. Although performance and computation quality are equally important, the principal aim of approximate computing is to reach fast, energy-efficient circuits and systems. Therefore, it would be pointless to have an approximate computing block with a negligible improvement over its precise counterpart. Unlike the approximate multiplier in [19], which fails to keep a balance between performance and computation quality (the performance is dramatically below average while having a negligible NED), the proposed multipliers in this paper reach an excellent compromise between computation and hardware qualities. Design #2 has the lowest PDAP, 86.34% lower than [19], with a reasonable MED and NED.

3. This paper shows that alongside MED, NED, and ER, the error pattern of an approximate multiplier can directly affect the results of an image processing application. This is why the generated images by the multiplier in [15] are utterly dark, whereas the ones by [21] keep the general structure of the images. Although both multipliers have almost the same MED and ER, their error patterns determine the eventual outcome. Besides, an approximate multiplier might not be appropriate for a specific application despite having a satisfactory outcome for another one.

Eventually, we suggest some future works:

1. The suitability of the proposed approximate multipliers in other applications can be investigated in future studies. We believe that they will be successful in most applications because of their moderately low MED and almost uniform error pattern.
2. This study shows that multicolumn inexact compressors have not received the attention they deserve. In the future, other multicolumn inexact compressors can be suggested for designing high-performance approximate multipliers.

This paper targets gate-level designs. It is possible to perform several low-level optimizations for the proposed compressors when it comes to transistor-level implementation.

# Appendix I

Table 6. Characteristics of the Derived Compressors from the Proposed Multicolumn 3,3:2 Inexact Compressor

| Compressor | Inner Structure | NED | $FOM_1$ | $FOM_2$ |
|---|---|---|---|---|
| 3,3:2 | | 0.08125 | 0.4820 | 4.7429 |
| 3,3:2 (Without $C_{in}$) | | 0.0555 | 0.3123 | 2.1521 |
| 3,2:2 (Without $C_{in}$) | | 0.03125 | 0.3744 | 1.6491 |
| 2,3:2 | | 0.10156 | 0.5779 | 3.8633 |
| 2,2:2 | | 0.07143 | 0.4617 | 2.1125 |
| 1,3:2 | | 0.13542 | 0.7640 | 3.3391 |
| 1,2:2 | | 0.1 | 0.7893 | 1.3055 |
| 1,2:2 (Without $C_{in}$) | | 0.0625 | 0.4599 | 0.2857 |

# Appendix II

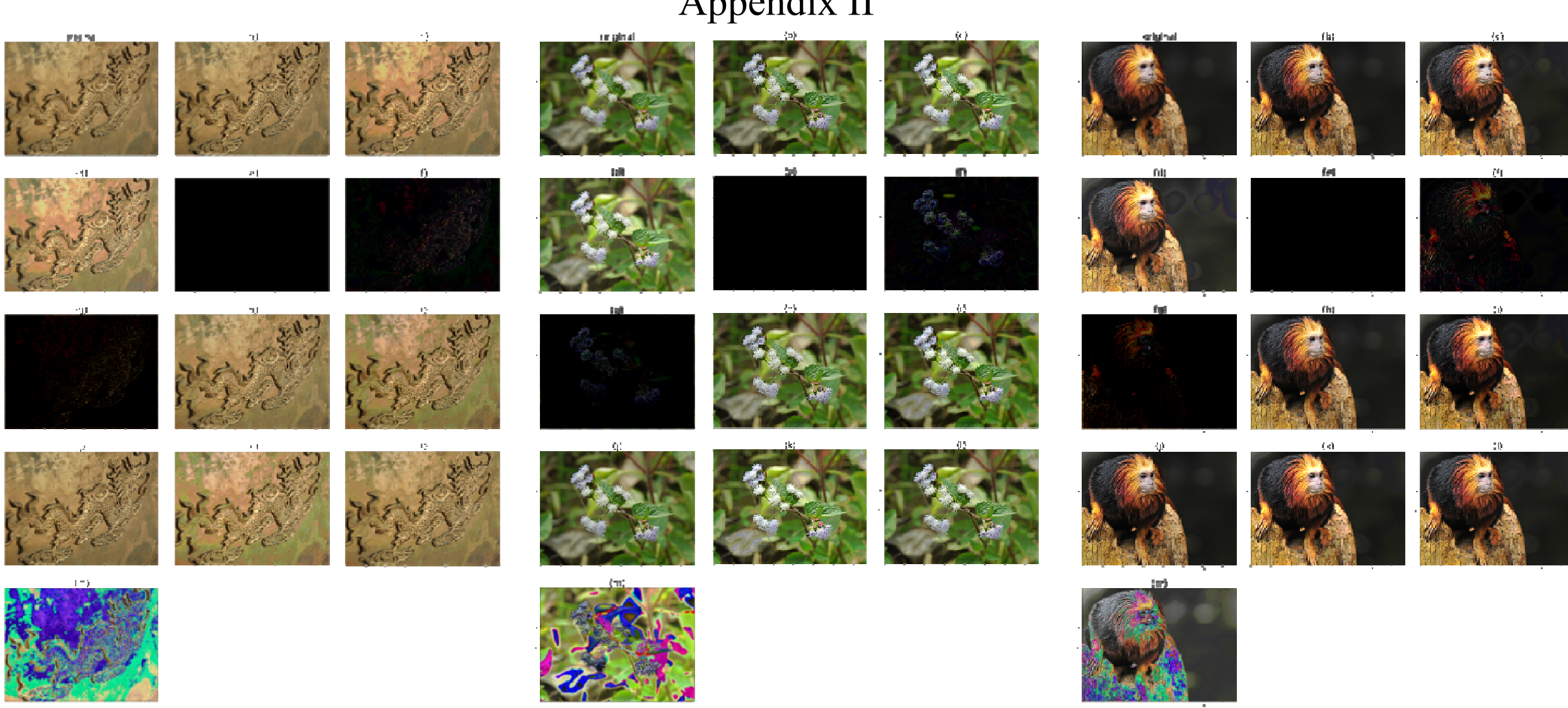

Fig. 14. Image of a dragon, (a) Original, and sharpened by (b) System multiplier, (c) Proposed (Design #1), (d) Proposed (Design #2), (e) [15], (f) [20], (g) [14], (h) [18], (i) [17], (j) [19], (k) [16], (l) [13], (m) [21].

Fig. 15. Image of a flower, (a) Original, and sharpened by (b) System multiplier, (c) Proposed (Design #1), (d) Proposed (Design #2), (e) [15], (f) [20], (g) [14], (h) [18], (i) [17], (j) [19], (k) [16], (l) [13], (m) [21].

Fig. 16. Image of a monkey, (a) Original, and sharpened by (b) System multiplier, (c) Proposed (Design #1), (d) Proposed (Design #2), (e) [15], (f) [20], (g) [14], (h) [18], (i) [17], (j) [19], (k) [16], (l) [13], (m) [21].

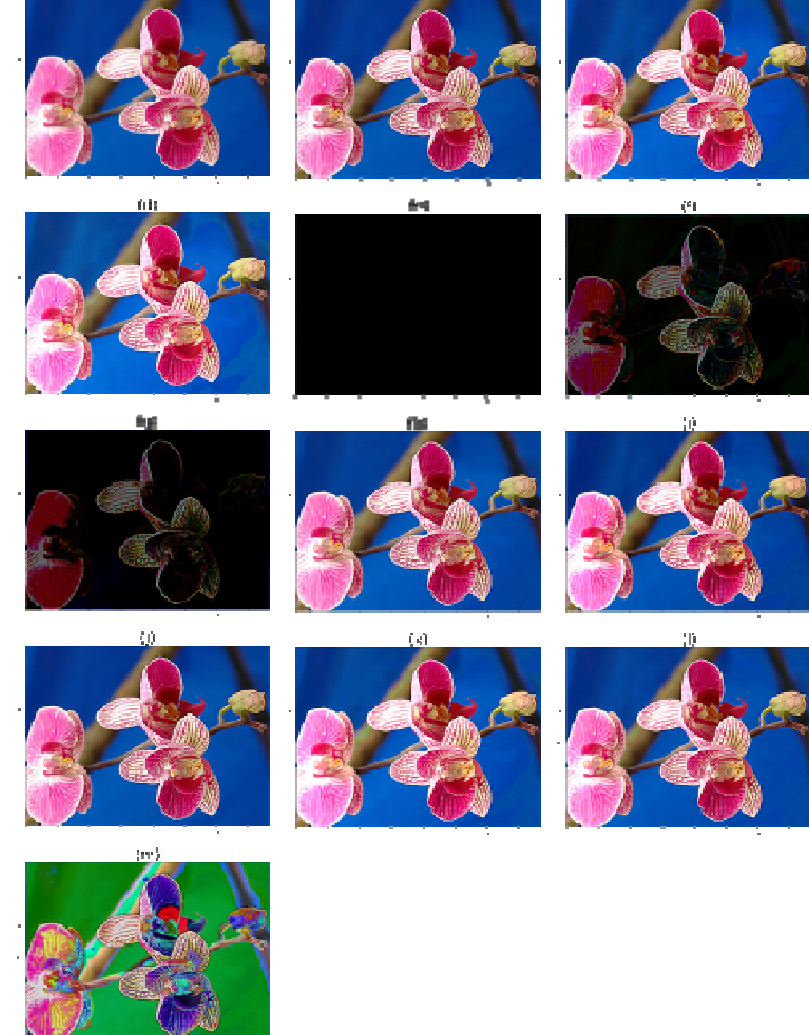

Fig. 17. Image of an orchid, (a) Original, and sharpened by (b) System multiplier, (c) Proposed (Design #1), (d) Proposed (Design #2), (e) [15], (f) [20], (g) [14], (h) [18], (i) [17], (j) [19], (k) [16], (l) [13], (m) [21].

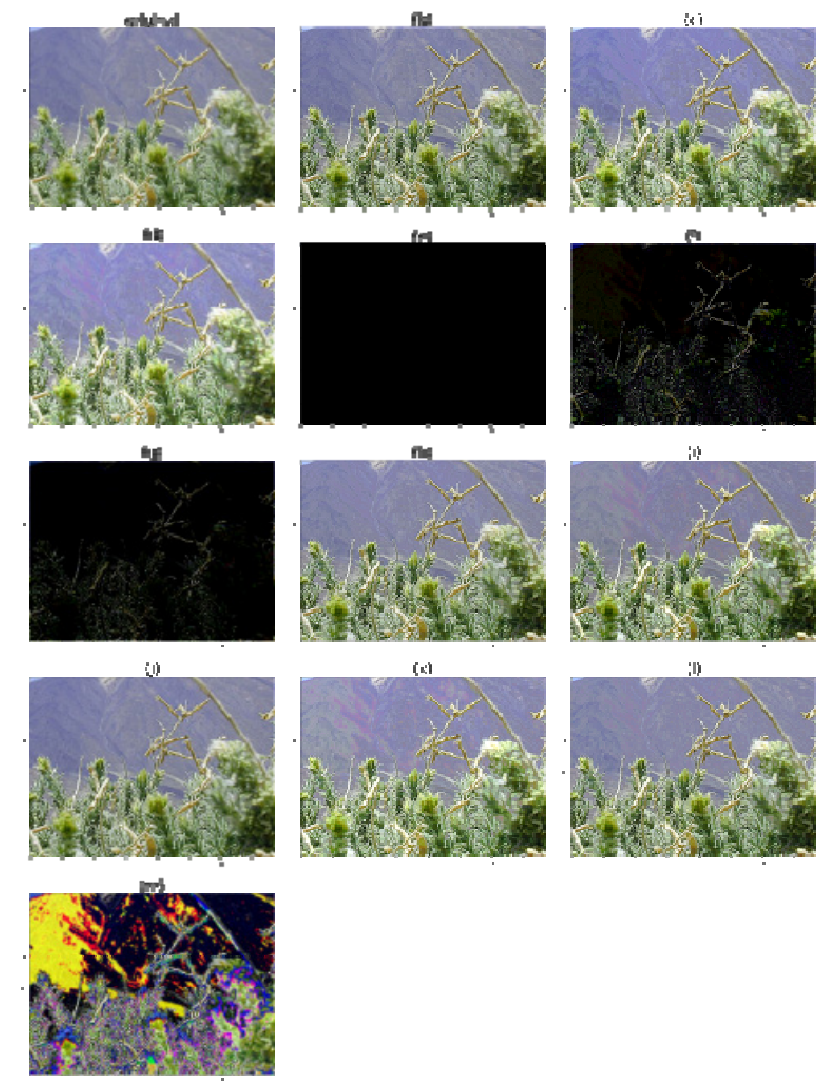

Fig. 18. Image of a peak, (a) Original, and sharpened by (b) System multiplier, (c) Proposed (Design #1), (d) Proposed (Design #2), (e) [15], (f) [20], (g) [14], (h) [18], (i) [17], (j) [19], (k) [16], (l) [13], (m) [21].

# References

[1] S. Mittal, "A survey of techniques for approximate computing," *ACM Computing Surveys,* vol. 48, no. 4, pp. 1-33, Mar. 2016.

[2] S. Venkataramani, V.K. Chippa, S.T. Chakradhar, K. Roy, and A. Raghunathan, "Quality programmable vector processors for approximate computing," *46th Annual IEEE/ACM Int. Symp. Microarchitecture, Davis, USA*, pp. 1-12, Dec. 2013.

[3] Q. Xu, T. Mytkowicz, and N.S. Kim, "Approximate computing: A survey," *IEEE Design & Test*, vol. 33, no. 1, pp. 8-22, Dec. 015.
[4] W. Liu, F. Lombardi, and M. Shulte, "A retrospective and prospective view of approximate computing," *Proc. of the IEEE*, vol. 108, no. 3 pp. 394-399, Mar. 2020.
[5] M. Samadi, J. Lee, D.A. Jamshidi, A. Hormati, and S. Mahlke, "Self-tuning approximation for graphics engines," *46th Annual IEEE/ACM Int. Symp. Microarchitecture, Davis, USA,* pp. 13-24, Dec. 2013.
[6] S. Sinha, and W. Zhang, "Low-power FPGA design using memoization-based approximate computing," *IEEE Trans. Very Large Scale Integration (VLSI) Systems*, vol. 24, no. 8 pp. 2665-2678, Feb. 2016.
[7] B. Garg, and G.K. Sharma, "A quality-aware Energy-scalable Gaussian Smoothing Filter for image processing applications," *Microprocessors and Microsystems*, vol. 45, pp. 1-9, Aug. 2016.
[8] G. Zervakis, K. Tsoumanis, S. Xydis, D. Soudris, and K. Pekmestzi, "Design-efficient approximate multiplication circuits through partial product perforation," *IEEE Trans. Very Large Scale Integration (VLSI) Systems*, vol. 24, no. 10, pp. 3105-3117, Mar. 2016.
[9] H. Uoosefian, K. Navi, R.F. Mirzaee, and M. Hosseinzadeh, "High-Performance CML approximate full adders for image processing application of laplace transform," *Circuit World*, vol. 46, no. 4, pp. 285-299, Apr. 2020.
[10] F. Ebrahimi-Azandaryani, O. Akbari, M. Kamal, A. Afzali-Kusha, and M. Pedram, "Block-based carry speculative approximate adder for energy-efficient applications," *IEEE Trans. Circuits and Systems II: Express Briefs*, vol. 67, no. 1, pp. 137-141, Feb. 2019.
[11] C. Dharmaraj, V. Vasudevan, and N. Chandrachoodan, "Optimization of signal processing applications using parameterized error models for approximate adders," *ACM Trans. Embedded Computing Systems*, vol. 20, no. 2, pp. 1-25, Jan. 2021.
[12] Y. Mannepalli, V.B. Korede, and M. Rao, "Novel approximate multiplier designs for edge detection application," *Great Lakes Symp. VLSI , Virtual, USA,* pp. 371-377, Jun. 2021.
[13] M. Ahmadinejad, M.H. Moaiyeri, and F. Sabetzadeh, "Energy and area efficient imprecise compressors for approximate multiplication at nanoscale," *AEU-Int. J. Electronics and Communications*, vol. *110*, p. 152859, Oct. 2019.
[14] F. Sabetzadeh, M.H. Moaiyeri, and M. Ahmadinejad, "A majority-based imprecise multiplier for ultra-efficient approximate image multiplication," *IEEE Trans. Circuits and Systems I: Regular Papers*, vol. 66, no. 11, pp. 4200-4208, Jun. 2019.
[15] A. Momeni, J. Han, P. Montuschi, and F. Lombardi, "Design and analysis of approximate compressors for multiplication," *IEEE Trans. Computers*, vol. 64, no. 4, pp. 984-994, Feb. 2014.
[16] S. Venkatachalam, and S.B. Ko, "Design of power and area efficient approximate multipliers," *IEEE Trans. Very Large Scale Integration (VLSI) Systems*, vol. 25, no. 5, pp. 1782-1786, Jan. 2017.
[17] R. Jothin, M.P. Mohamed, and C. Vasanthanayaki, "High performance compact energy efficient error tolerant adders and multipliers for 16-bit image processing applications," *Microprocessors and Microsystems*, vol. 78, p. 103237, Oct. 2020.
[18] X. Yi, H. Pei, Z. Zhang, H. Zhou, and Y. He, "Design of an energy-efficient approximate compressor for error-resilient multiplications," *IEEE Int. Symp. Circuits and Systems, Sapporo, Japan,* pp. 1-5, May 2019.
[19] A.G.M. Strollo, E. Napoli, D. De Caro, N. Petra, and G. Di Meo, "Comparison and extension of approximate 4-2 compressors for low-power approximate multipliers," *IEEE Trans. Circuits and Systems I: Regular Papers*, vol. 67, no. 9, pp. 3021-3034, May 2020.
[20] K.M. Reddy, M.H. Vasantha, Y.N. Kumar, and D. Dwivedi, "Design and analysis of multiplier using approximate 4-2 compressor," *AEU-Int. J. Electronics and Communications*, vol. 107, pp. 89-97, Jul. 2019.
[21] M. Taheri, A. Arasteh, S. Mohammadyan, A. Panahi, and K. Navi, "A novel majority based imprecise 4: 2 compressor with respect to the current and future VLSI industry," *Microprocessors and Microsystems*, vol. 73, p. 102962, Mar. 2020.
[22] M. Imran, M. Rashid, A.R. Jafri, and M. Kashif, "Throughput/area optimised pipelined architecture for elliptic curve crypto processor," *IET Computers & Digital Techniques*, vol. 13, no. 5, pp. 361-368, Sep. 2019.
[23] D.M. Ellaithy, M.A. El-Moursy, A. Zaki, and A. Zekry, "Dual-Channel Multiplier for Piecewise-Polynomial Function Evaluation for Low-Power 3-D Graphics," *IEEE Trans. Very Large Scale Integration (VLSI) Systems*, vol. 27, no. 4, pp. 790-798, Jan. 2019.
[24] S.R. Faraji, and K. Bazargan, "Hybrid binary-unary truncated multiplication for DSP Applications on FPGAs," *IEEE/ACM Int. Conf. Computer Aided Design, San Diego, USA,* pp. 1-9, Nov. 2020.
[25] A. Wang, G.A. Jullien, and W.C. Miller, "A new design technique for column compression multipliers," *IEEE Trans. Computers*, vol. 44, no. 8, pp. 962-970, Aug. 1995.

[26] C.S. Wallace, “A suggestion for a fast multiplier,” *IEEE Trans. Electronic Computers*, vol. EC-13, no. 1, pp. 14-17, Feb. 1964.

[27] L. Dadda, “Some schemes for parallel multipliers,” *Alta Frequenza*, vol. 34, no. 5, pp. 349-356, 1965.

[28] S. Mehrabi, R.F. Mirzaee, S. Zamanzadeh, and A. Jamalian, “Multiplication With m:2 and m:3 Compressors—A Comparative Review,” *Canadian J. Electrical and Computer Engineering*, vol. 40, no. 4, pp. 303-313, Dec. 2017.

[29] M. Rafiee, Y. Sadeghi, N. Shiri, and A. Sadeghi, “An approximate CNTFET 4:2 compressor based on gate diffusion input and dynamic threshold,” *Electronics Lett.*, Early View, May 2021.

[30] R. Marimuthu, Y.E. Rezinold, and P.S. Mallick, “Design and analysis of multiplier using approximate 15-4 compressor,” *IEEE Access*, vol. 5, pp. 1027-1036, Dec. 2016.

[31] S. Shirzadeh, and B. Forouzandeh, “High accurate multipliers using new set of approximate compressors,” *AEU-Int. J. Electronics and Communications*, In Press, Apr. 2021.

[32] B. Fang, H. Liang, D. Xu, M. Yi, Y. Sheng, C. Jiang, Z. Huang, and Y. Lu, “Approximate multipliers based on a novel unbiased approximate 4-2 compressor,” *Integration*., vol. 81, pp. 17-24, Nov. 2021.

[33] M. Ahmadinejad, and M.H. Moaiyeri, “Energy-and quality-efficient approximate multipliers for neural network and image processing applications,” *IEEE Trans. Emerging Topics in Computing*, Early Access, Apr. 2021.

[34] H. Pei, X. Yi, H. Zhou, and Y. He, “Design of ultra-low power consumption approximate 4–2 compressors based on the compensation characteristic,” *IEEE Trans. Circuits and Systems II: Express Briefs*, vol. 68, no. 1, pp. 461-465, Jun. 2020.

[35] A.K. Uppugunduru, and S.E. Ahmed, “Hardware-efficient approximate multiplier architectures for media processing applications,” *Circuit World*, Early Cite, Mar. 2021.

[36] V.K. Odugu, “An efficient VLSI architecture of 2-D finite impulse response filter using enhanced approximate compressor circuits,” *Int. J. Circuit Theory and Applications*, vol. 49, no. 11, pp. 3653-3668, Nov. 2021.

[37] Y. Zhang, X. Chen, C. He, and G. Xie, “Energy-efficient multipliers using imprecise compressors for image multiplication,” *Int. J. Circuit Theory and Applications*, vol. 50, no. 11, pp. 3875-3890, Nov. 2022.

[38] D. Wang, T. Gao, and Y. Zhang, “Image sharpening detection based on difference sets,” *IEEE Access*, vol. 8, pp. 51431-51445, Mar. 2020.

[39] W. Ma, and S. Li, “A new high compression compressor for large multiplier,” *9th Int. Conf. Solid-State and Integrated-Circuit Technology, Beijing, China*, pp. 1877-1880, Oct. 2008.

[40] M.S. Rashid, *Investigation of a method to identify energy efficient organization of compression unit in parallel tree multipliers*, Master's Thesis, Middle East Technical University, 2018.

[41] C.T. Vu, T.D. Phan, and D.M. Chandler, “$S_3$: A spectral and spatial measure of local perceived sharpness in natural images,” *IEEE Trans. Image Processing*, vol. 21, no. 3, pp. 934-945, Sep. 2011.

[42] Local Image Sharpness Database, available at: http://vision.eng.shizuoka.ac.jp/pluginfile.php/71/mod_page/content/9/sharpnessmapdb.zip

[43] M.S. Lau, K.V. Ling, and Y.C. Chu, “Energy-aware probabilistic multiplier: design and analysis,” *Int. Conf. Compilers, Architecture, and Synthesis for Embedded Systems, Grenoble, France*, pp. 281-290, Oct. 2009.

[44] N. Maheshwari, Z. Yang, J. Han, and F. Lombardi, “A design approach for compressor based approximate multipliers,” *28th Int. Conf. VLSI Design, Bangalore, India,* pp. 209-214, Jan. 2015.

[45] L. Shapiro, and G. Stockman. *Computer Vision*, CA, USA: Prentice Hall. 2001, p. 137.